\documentclass{PoS}

\usepackage{graphicx}
\usepackage{amsmath}
\usepackage{amssymb}
\usepackage{slashed}

\title{From lattice QCD to predictions of scattering phase shift at the physical point}

\ShortTitle{From lattice QCD to predictions of scattering phase shift at the physical point}

\author{\speaker{Xiao-Yu Guo}\\
        Beijing University of Technology,\\
        Beijing 100124, China\\
        GSI Helmholtzzentrum f\"ur Schwerionenforschung GmbH, \\Planckstra\ss e 1, 64291 Darmstadt, Germany\\
        E-mail: \email{x.guo@gsi.de}} 
\author{Yonggoo Heo\\
        Suranaree University of Technology, Nakhon Ratchasima, 30000, Thailand\\
        E-mail: \email{y.heo@g.sut.ac.th}}  
\author{Matthias F.M. Lutz\\
        GSI Helmholtzzentrum f\"ur Schwerionenforschung GmbH, \\Planckstra\ss e 1, 64291 Darmstadt, Germany\\
        E-mail: \email{m.lutz@gsi.de}}

\abstract{
The Hadron Spectrum Collaboration (HSC) presented new results on
two of their ensembles for s-wave scattering phase shifts
in the open-charm sector of QCD. For such ensembles we have made
predictions that are based on the chiral Lagrangian that were published two years ago. In this talk we confront our phase shifts with those of HSC. 
A remarkably consistent picture emerges. In particular there is mounting
evidence for the existence of a flavor-sextet state in the $D \pi$
and $D^*\pi$ channels, that show a striking quark-mass dependence.
}

\FullConference{%
 The 38th International Symposium on Lattice Field Theory, LATTICE2021\\
  26th-30th July, 2021\\
  Zoom/Gather@Massachusetts Institute of Technology
}

\begin{document}

\maketitle

\section{Introduction}

In our previous study \cite{Guo:2018kno} the available lattice QCD data set in the open-charm meson sector 
was confronted with a novel SU(3) chiral perturbation theory (ChPT) approach. For the first time the $D$- and $D^*$-meson masses 
were considered simultaneously with the first few scattering phase shifts. This is important since some of the low-energy constants in the chiral Lagrangian contribute 
to both, the meson masses but also to the scattering processes. The convergence of conventional SU(3) 
ChPT was  much improved by insisting on the use of on-shell masses rather than chirally expanded masses in all considered loop contributions \cite{Lutz:2018cqo}. A global fit to the lattice data was performed. Low-energy constants (LEC) at leading and sub-leading orders were adjusted such that the lattice data at unphysical quark masses are recovered. The lattice data consisted of the masses of open-charm meson ground-states from 5 different lattice groups including ETMC, HPQCD, HSC, LHPC and PACS-CS \cite{Kalinowski:2015bwa,Na:2012iu,Moir:2016srx,Liu:2012zya,Mohler:2011ke,Lang:2014yfa}. 
In addition, we considered the scattering lengths on some LHPC ensembles
as given in \cite{Liu:2012zya} and the the  s-wave  $D \pi$-phase shifts on an HSC ensemble \cite{Moir:2016srx}. 

In application of our global fit, we were able to make specific predictions at physical light-quark masses.
Most striking are our results on the coupled-channel $D \pi$ s-wave system, for which HSC provided first results on an ensemble with a pion mass  $m_\pi \simeq  380$ MeV. 
The phase shifts are characterized by a striking quark-mass dependence. Such a system with  $J^P =0^+$ quantum numbers has been of particular interest in studies of $D$-meson resonances as a two-pole structure has been predicted \cite{Kolomeitsev:2003ac,Hofmann:2003je,Altenbuchinger:2013vwa,Du:2017zvv,Albaladejo:2016lbb,Guo:2018tjx,Du:2020pui,Gregory:2021rgy}. Analogous results are expected for the $D^* \pi$ system with $J^P =1^+$\cite{Guo:2018gyd}. 
We confirm the two-pole structure in the amplitudes. One pole belongs to the flavor anti-triplet, the other belongs to the flavor sextet. 

In order to prepare a further test bed for this scenario we published with \cite{Guo:2018kno} a dedicated study on a second HSC ensemble characterized by a significantly smaller pion mass $m_\pi\simeq 220$ MeV.  
To this end, we calculated different D-meson scattering phase shifts not only for physical light-quark masses, but also for unphysical quark masses  with  $m_\pi\simeq 220$ MeV and $380$ MeV as implied by two HSC ensembles. Recently, HSC has unleashed a series of works presenting various open-charm meson scattering phase shifts on those two ensembles \cite{Cheung:2020mql,Gayer:2021xzv}. 

In this contribution to LATTICE2021 we show a first comparison of our predictions with those recent results by HSC. For the details of our global fit we refer to the original work \cite{Guo:2018kno}.

\section{Comparison with lattice results}

\begin{figure}[t]
\center{
\includegraphics[keepaspectratio,height=0.38\textwidth]{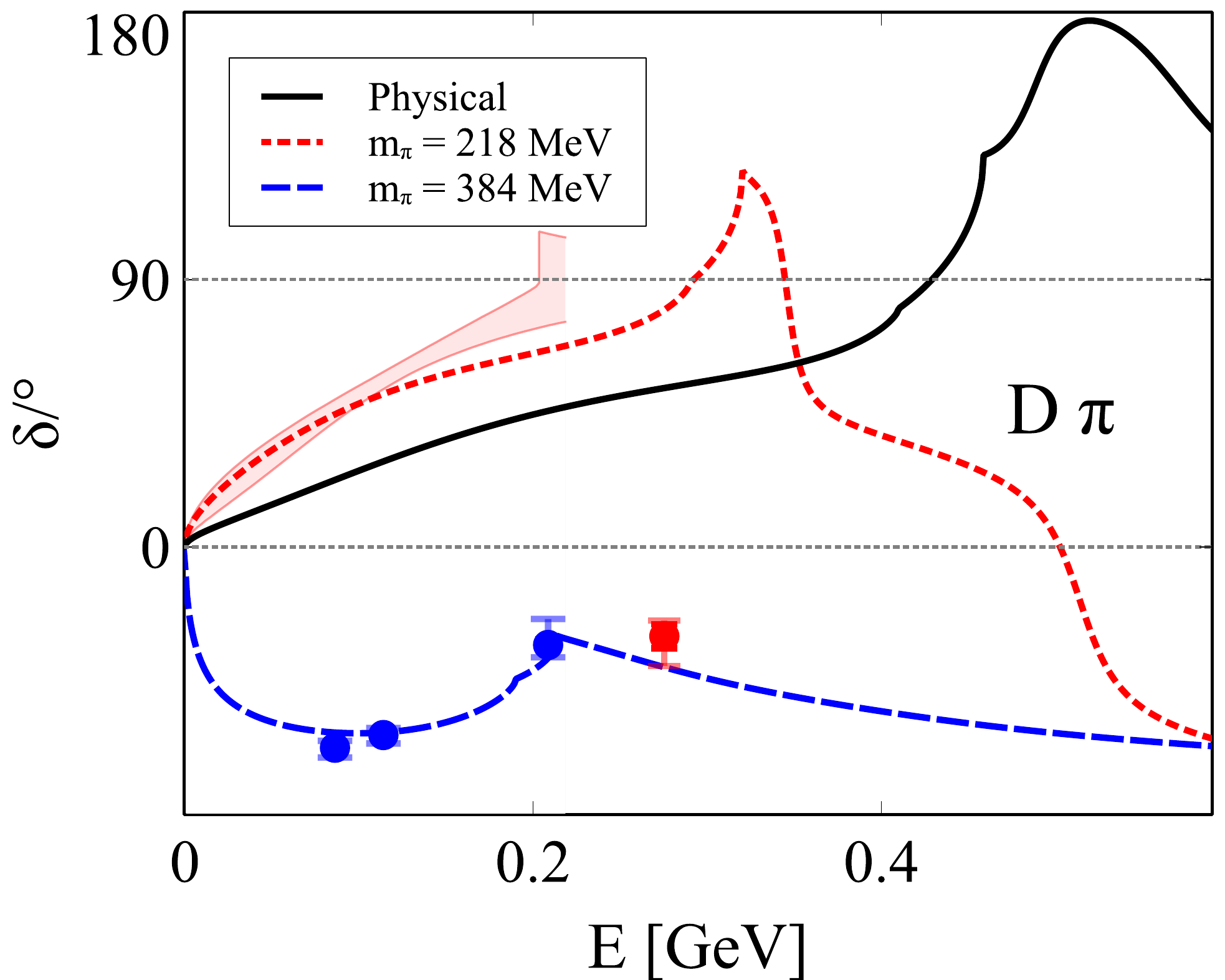}
\includegraphics[keepaspectratio,height=0.38\textwidth]{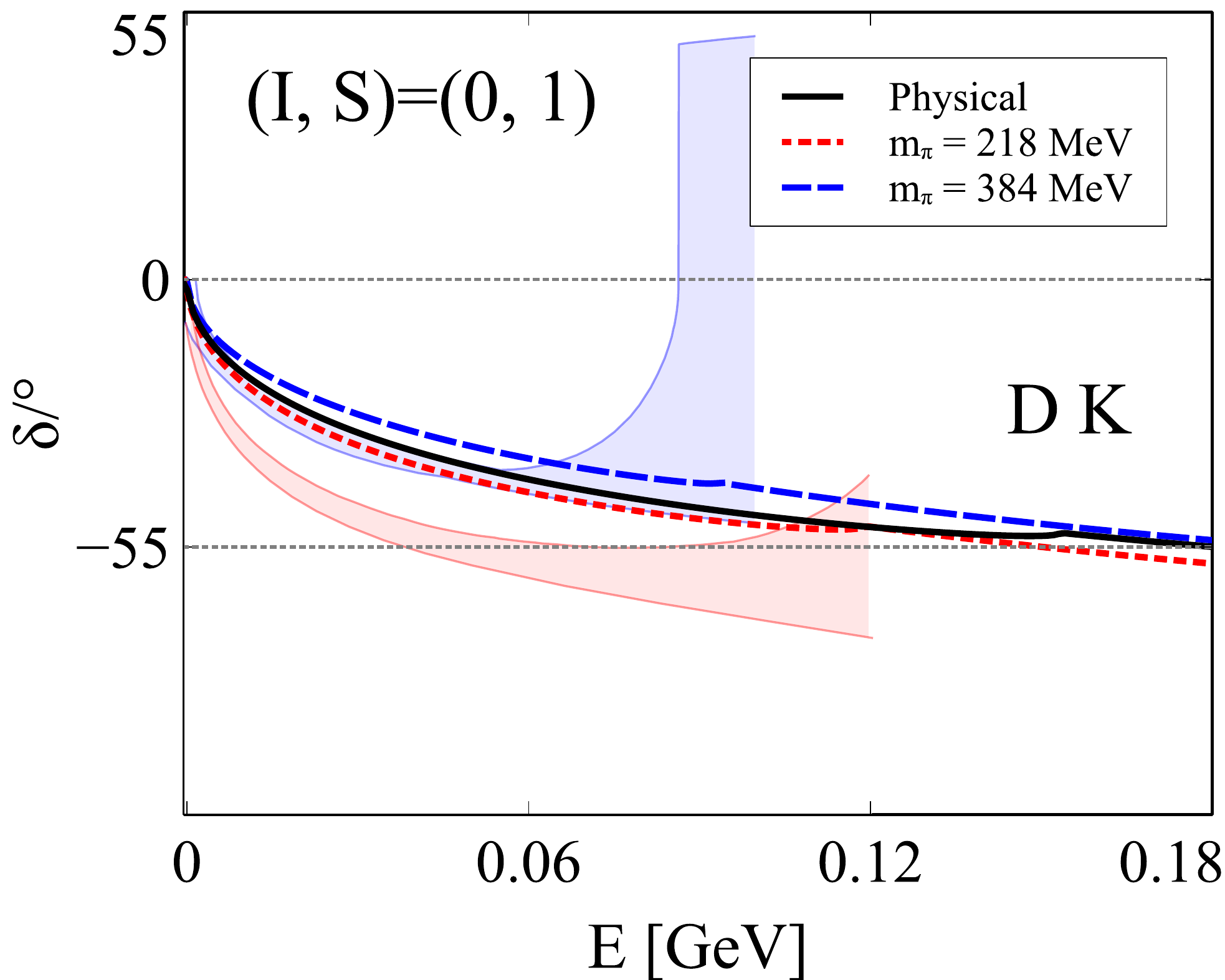}
}
\vskip-0.2cm
\caption{\label{fig:1} The  s-wave $D \pi$ with $(I,S) = (1/2,0)$ (left panel) and s-wave $DK$ with $(I,S) = (0,1)$ (right panel) phase shifts at different light-quark masses. The black-solid, red-dotted and blue-dashed curves represent our predictions at physical quark masses and on two HSC ensembles with $m_\pi = 218$ MeV and $m_\pi = 384$ MeV respectively. The blue (red) bands are the lattice data on the $m_\pi = 384$  MeV ($m_\pi = 218$ MeV) ensemble from Refs.\,\cite{Cheung:2020mql,Gayer:2021xzv}. The blue (red) points in the left panel are the lattice data on the $m_\pi = 384$ MeV ensemble included (excluded) in our fit. }
\end{figure}

We first focus on the scattering process with isospin-strangeness $(I,S) = (1/2,0)$. It involves three distinct 2-body final states $D \pi$, $D\eta$ and $D_s\bar K$. 
The lattice results on the $m_\pi \simeq 380$ MeV ensemble were published a few years ago \cite{Moir:2016srx}. The blue and red points in  Fig.\,\ref{fig:1} reflect such results, where we used a binning of the published phase-shifts into suitable single-energy data points.  
Such blue phase shifts had been incorporated in our global fit. 
They should be compared with the dashed-blue line from our best fit.
Based on the LEC obtained, we calculated phase shifts for the  $m_\pi \simeq 220$ MeV ensemble and the physical case. The pion mass in physical units on that second ensemble is distinct from the one given in \cite{Cheung:2020mql,Gayer:2021xzv}. This is so since we used a mass-independent scale setting in \cite{Guo:2018kno}.
The additional $D \pi$ phase-shift lines are shown in the left panel of Fig.\,\ref{fig:1}. They are plotted as functions of the kinetic energy in the center-of-mass frame $E\equiv \sqrt{s} - \sqrt{s}_{\rm thr}$ with $\sqrt{s}_{\rm thr} = M_D+m_\pi$.
The black-solid line is our result for the phase shift in the physical case. A striking quark-mass dependence was predicted.  

Recently, HSC has provided phase shifts on their second ensemble. 
The result covers the energy range from the $D \pi$ threshold to the three-body threshold $D\pi \pi$.
The lattice results are plotted in the left panel of Fig.\ref{fig:1} as a red band. 
The width of the band indicates the error size, which is estimated according to the left panel of Fig.\,10 in Ref.\,\cite{Gayer:2021xzv}. 
We translate the lattice result into physical unit, by matching the energy interval between the $D \pi$ threshold $\sqrt{s}_{\rm thr}$ and the $D \pi \pi$ threshold. 
Comparing the lattice result to our prediction (red-dotted line), we observe that they roughly agree especially near the $D \pi$ threshold. As the energy increases, the lattice result of the phase shift rises with a similar rate as our prediction. 

Two poles on unphysical Riemann sheets are shaping the phase-shift lines. The first pole is below the $D\eta$ threshold quite away from the real axis and the second one around the $D\eta$ threshold closer to the real axis. The phase shift escalates as the energy increases above the $D \pi$ threshold passing through 90 degrees. This feature is driven by the second pole in the complex plane. 
This can be partially confronted with the lattice results. HSC reports one pole on the second Riemann sheet.
Using our scale setting, the pole mass is $69(\pm 61)$ MeV above the $D \pi$ threshold. It comes with a  width of $\Gamma = 389(\pm 206)$ MeV. 
On the other hand, our favorite set of LEC suggests the pole to lie around $2$ MeV above the threshold, with a width $\Gamma = 257(_{+71}^{-55})$ MeV. The uncertainty in our real part is quite large. While it is hard to move the pole further to the right, it may go as much as $130 $ MeV further left in the complex plane. 
The results from the two approaches appear compatible qualitatively. Our prediction for the second pole, 
which is associated with a flavor sextet, comes at $334(_{-34}^{+49})$ MeV above the $D \pi$ threshold 
with a width of $\Gamma =  193(_{+20}^{-44})$ MeV. At the physical point we recall our published values 
$433(_{-32}^{+41})$ MeV with $\Gamma  =185(_{+6}^{-14})$ MeV. Corresponding values from HSC are so far not available.  

We continue with the $(I,S) = (0,1)$ channel. Here, two final states $DK$ and $D_s\eta$ play a role.  
In our previous work, we predicted the $DK$ phase shift at both physical quark masses and unphysical quark masses with $m_\pi \simeq 220$ MeV and $380$ MeV corresponding to the two HSC ensembles already encountered previously. With the recent HSC work \cite{Cheung:2020mql}, phase shifts on two ensembles became available. A comparison of our prediction against the lattice results is provided in the right panel of Fig.\,\ref{fig:1}.   
In the physical case, the scattering amplitude shows a pole below the $DK$ threshold, which has been recognized as the $D^*_{s0}(2317)$  \cite{Kolomeitsev:2003ac,Lutz:2007sk,Guo:2006fu,Altenbuchinger:2013gaa,Guo:2015dha,Du:2017ttu,Wu:2019vsy}. Our prediction shows that this pole is quite stable against a variation of the light-quark masses. In the right panel of Fig.\,\ref{fig:1}, both  cases with $m_\pi \simeq 220$ MeV and 380 MeV, contain a pole below the $DK$ threshold. Such a behavior leads to a decline of the phase shift, as indicated in the figure by the black-solid, red-dotted and blue-dashed lines. 
This feature is confirmed by the recent HSC lattice calculation. The lattice results are plotted in red and blue bands, with an error estimation based on Fig.\,10 of Ref.\cite{Cheung:2020mql}.
They imply a pole located at energies $23(3)$ MeV and $56(3)$ MeV below the $DK$ threshold. 
We obtain poles at $66(_{-32}^{+34})$ MeV and $127(_{-41}^{+40})$ MeV below the threshold in comparison. Since the pole is close to the $D K$ threshold, the shape of the phase-shift line is heavily affected by the position of the pole. 
Although the predicted magnitude of the phase shifts are in qualitative agreement with the lattice results, there remain significant discrepancies in particular for the $m_\pi \simeq 220$ MeV ensemble. This suggests further studies that should incorporate such lattice data in a global fit.

\begin{figure}[b]
\center{
\includegraphics[keepaspectratio,height=0.38\textwidth]{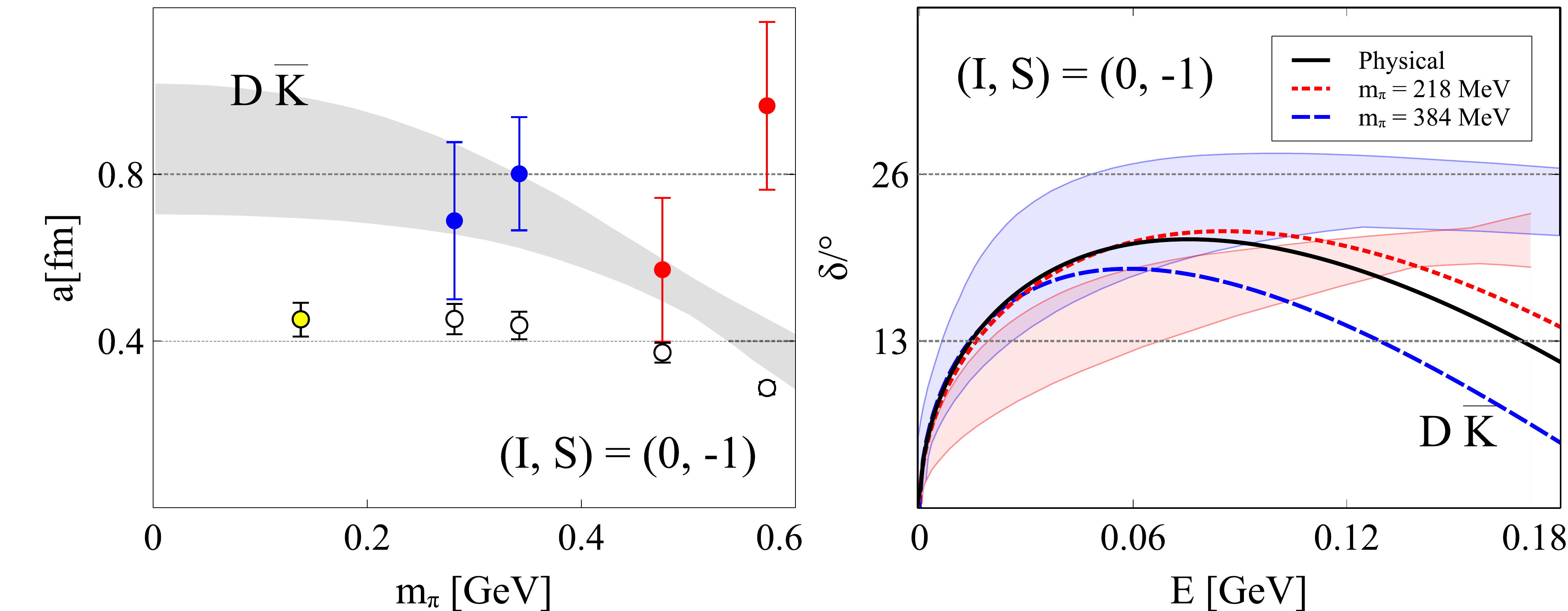}
}
\vskip-0.2cm
\caption{\label{fig:2} The $(I,S) = (0,-1)$ $D\bar K$ scattering lengths (left panel) and phase shifts (right panel) at different light-quark masses. For the left panel, the blue (red) points represent the LHPC data points included (excluded) in our fits. The scattering lengths from our fit  are in empty points, whereas the scattering length extrapolated to the physical case is the yellow point. For comparison, the scattering lengths obtained by Liu et. al. \cite{Liu:2012zya} are recalled in a grey band. For the right panel, the black-solid, red-dotted and blue-dashed curves represent our predictions at physical quark masses and on two HSC ensembles with $m_\pi = 218$ MeV and $m_\pi = 384$ MeV respectively. The blue (red) bands are the lattice data on the $m_\pi = 384$ MeV ($m_\pi = 218$ MeV) ensemble from Ref.\,\cite{Cheung:2020mql}.}
\end{figure}

\begin{figure}[t]
\center{
\includegraphics[keepaspectratio,height=0.38\textwidth]{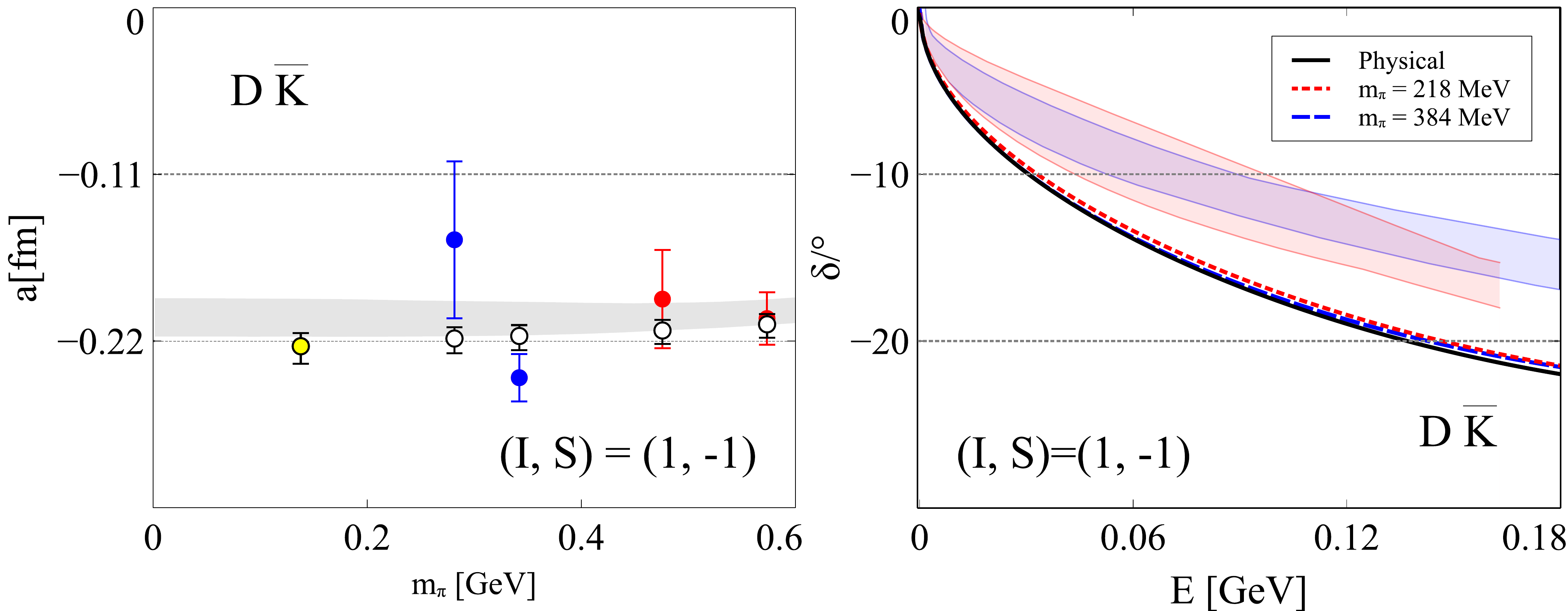}
}
\vskip-0.2cm
\caption{\label{fig:3} The $(I,S) = (1,-1)$ $D\bar K$ scattering lengths (left panel) and phase shifts (right panel) at different light-quark masses. The color coding is the same as in Fig.\ref{fig:2}. }
\end{figure} 

We emphasize that the lattice results driving us to our predictions do not only include the HSC $ D \pi$ phase shifts but also some scattering lengths on LHPC ensembles from Ref.\,\cite{Liu:2012zya}. Given our findings it is an important issue to explore the degree of compatibility of the distinct data sets. This issue emerges in the $D\bar K$ scattering process, for which there are lattice results from both types of ensembles available. Two $D\bar K$ channels with $(I,S) = (0,-1)$ or $(I,S) = (1,-1)$ are possible. 
We first look into the $(I,S) = (0,-1)$ system. From our LEC, we predicted the $D\bar K$ phase shifts for both ensembles. The phase shifts are shown in the right panel of Fig. \ref{fig:2}. The red-dotted line gives the result for the $m_\pi\simeq 220$ MeV ensemble, the blue-dashed line for the $m_\pi \simeq 380$ MeV ensemble. The two lines should be compared with corresponding HSC bands, taken from  Fig.\,15 of \cite{Cheung:2020mql}. Although our predictions qualitatively agree with the lattice result, there are again  quantitative discrepancies. Here it is possible to confront such a pattern with corresponding results 
on LHPC ensembles. In the left-hand panel of Fig.\,\ref{fig:2} they are recalled and plotted as a function of the pion mass.  Most striking are data on the HSC ensemble at $m_\pi \simeq 220$ MeV: the results suggest that our LEC imply a significant overestimate of the scattering length at the physical point. 
Here we note, that due to the global nature of our fit, our scattering length was already largely  pulled down to smaller values, against the request of the lattice scattering lengths. 
For comparison, we recall the scattering lengths from the competing approach \cite{Liu:2012zya}. They are shown by the grey band in the left-hand panel of Fig.\,\ref{fig:2}.  At the physical point their scattering length was predicted to be almost twice as large as ours. This approach puts more emphasis on the LHPC ensembles. Constraints from ETMC, HPQCD, PACS and HSC results on open-charm meson masses were not considered. This feature may hint at some tension amongst the two complementary lattice approaches.

We close our discussion with the  $(I,S) = (1,-1)$ system, for which we collect various results in Fig. \ref{fig:3}. 
In the right-hand panel of the figure, our predictions are plotted against the HSC error bands as given by Fig.\,15 of \cite{Cheung:2020mql}.
Our prediction of a rather mild pion-mass dependence of the phase shift is confirmed by the lattice results. However, there is a sizeable gap between our and the lattice phase shifts. Close to threshold, that would imply an overestimate of the magnitude of the scattering length. Given this observation it is enlightening to look into the left-hand panel of the figure, in which our results are confronted with scattering lengths from the LHPC ensembles. 
We considered only the two blue-colored data points. Here the second point from the left prevented the fit from leading to a larger scattering length as it seems
mandated by the HSC results. 
Similar results from \cite{Liu:2012zya} are shown by the grey band in the left-panel of Fig.\,\ref{fig:3}.

\section{Summary}

This report offers a critical discussion of predictions we have made about two years ago for 
open-charm meson systems. The predictions rest on the lattice data set available 
at the time and 
are applications of the chiral SU(3) Lagrangian. A quantitative comparison with phase shifts recently released by HSC on two ensembles with a light and a heavy pion mass of about 220 MeV and 380 MeV was 
discussed in depth. Most striking we find our successful prediction of the quite large pion-mass dependence of the $D\pi$ $s$-wave scattering phase shift. This result gives strong support for the existence of an exotic flavor-sextet state in the open-charm meson sector of QCD. Moreover we uncovered a significant tension in the $D\bar K$ elastic $s$-wave scattering phase shifts, as constrained by lattice studies on HSC and LHPC ensembles. Further studies are desired to address this issue in more detail.

\bibliographystyle{JHEP}
\bibliography{thesis}

\end{document}